\documentclass[doublecol]{epl2} 
\usepackage{bm}
\usepackage{amsmath}
\usepackage{amsfonts}

\title{Electrochemical transport in Dirac nodal-line semimetals}

\author{R. Flores-Calder\'{o}n\inst{1,2} \and Leonardo Medel\inst{3} \and A. Mart\'{i}n-Ruiz\inst{3}}

\institute{                    
  \inst{1} Max Planck Institute for the Physics of Complex Systems, Nöthnitzer Str. 38, 01187 Dresden, Germany \\
  \inst{2} Max Planck Institute for Chemical Physics of Solids, Nöthnitzer Str. 40, 01187 Dresden, Germany\\
  \inst{3} Instituto de Ciencias Nucleares, Universidad Nacional Aut\'{o}noma de M\'{e}xico, 04510 Ciudad de M\'{e}xico, M\'{e}xico
  }

\pacs{05.20.Dd}{Kinetic theory}
\pacs{03.65.Vf}{Phases: geometric; dynamic or topological}
\pacs{65.40.gk}{Electrochemical properties}
 
\abstract{
Nodal-line semimetals are topological phases where the conduction and the valence bands cross each other along one-dimensional lines in the Brillouin zone, which are symmetry protected by either spatial symmetries or time-reversal symmetry. In particular, nodal lines  protected by the combined $\mathcal{PT}$ symmetry exhibits the parity anomaly of 2D Dirac fermions. In this Letter, we study the electrochemical transport in a $\mathcal{PT}$-symmetric Dirac nodal line semimetals by using the semiclassical Boltzmann equation approach. We derive a general formula for the topological current that includes both the Berry curvature and the orbital magnetic moment. We first evaluate the electrochemical current by introducing a small $\mathcal{PT}$-breaking mass term (which could be induced by inversion-breaking uniaxial strain, pressure, or an external electric field) and apply it to the hexagonal pnictide CaAgP. The electrochemical current vanishes in the zero-mass limit. Introducing a tilting term that does not spoil $\mathcal{PT}$ symmetry that protects the nodal ring, we obtain a finite electrochemical current in the zero-mass limit, which can be regarded as a direct consequence of the parity anomaly. We show that the parity anomaly induced electrochemical transport is also present at nonzero temperatures.}

\begin{document}

\maketitle

\section{Introduction}

Topological semimetals are new states of matter characterized by momentum space invariants over the Fermi surface \cite{Armitage}, rather than in the Brillouin zone as in topological insulators \cite{RevModPhys.83.1057}. They exhibit band crossings at point or lines in the Brillouin zone, and the band degeneracy at the contact is protected by symmetries such as crystalline and time-reversal \cite{10-fold-way}. Specifically, in Weyl semimetals (WSMs), the bands closest to the Fermi level cross each other at a discrete set of points \cite{Armitage}, while intersect along closed loops for Dirac nodal-line semimetals (DNLSMs) \cite{PhysRevB.92.081201, doi:10.1080/23746149.2022.2065216, Fang_2016}.

DNLSMs can be protected by time-reversal and non-symmorphic symmetries, by mirror planes, or by the combined symmetry of inversion $\mathcal{P}$ and time-reversal $\mathcal{T}$ \cite{Fang_2016}, such as the hexagonal pnictides CaAgP, CaAgAs and Ca$_3$P$_2$. In the latter case, they are endowed with a topological $\mathbb{Z}_2$ invariant, which is given by the phase winding of the Bloch states around the nodal loop \cite{PhysRevB.84.235126, PhysRevB.97.161113,PhysRevB.97.165104, PhysRevB.95.075138}. It is also known that, in the absence of spin-orbit coupling, the combined parity and time-reversal symmetry $\mathcal{PT}$ provides the symmetry protection of the nodal line \cite{PhysRevB.92.081201}. Therefore, any perturbation that preserves $\mathcal{PT}$ symmetry cannot remove the band crossing and open a gap. Anomalous transport induced by quantum anomalies has attracted great interest in the last decade. For example, the (3+1)-dimensional chiral anomaly in WSMs, which is intimately linked to the nontrivial topology of the band structure, leads to a negative longitudinal magnetoresistence \cite{PhysRevB.88.104412, PhysRevLett.113.247203, Negative_magnetoresistance} and the planar Hall effect \cite{PhysRevLett.119.176804, PhysRevB.102.121105, PhysRevB.99.115121}. In the case of $\mathcal{PT}$-symmetric DNLSMs, the (2+1)-dimensional parity anomaly has shown to determine linear and nonlinear topological currents. The parity anomaly is also realized in graphene \cite{PhysRevLett.61.2015} and leads to an interesting anomalous transport: the valley Hall effect \cite{PhysRevLett.99.236809}, which has attracted great attention due to its potential applications in valleytronic devices \cite{doi.org/10.1038/nphys547}. In the case of DNLSMs, which are (3+1)-dimensional systems, the low energy fermionic excitations can be described by a family of (2+1)-dimensional quantum field theories with a parity anomaly \cite{PhysRevB.97.161113, PhysRevB.97.165104}. In short, one can consider the DNLSM as a collection of graphene sheets. Therefore, although the parity anomaly occurs only in (2+1) dimensions, it also appears in DNLSMs, which are (3+1)-dimensional systems. Interestingly, since graphene is $\mathcal{PT}$ symmetric and its Dirac points have codimension $d_{c}=1$, it belongs to the same topological classification of DNLSMs \cite{PhysRevLett.116.156402}.

So far, magnetotransport experiments are the only probe of the topological $\mathbb{Z}_2$ invariant of DNLSMs \cite{PhysRevLett.117.016602, doi:10.1126/sciadv.1601742, PhysRevB.95.245113}. Therefore it is natural to ask us for other anomalous transport properties behind the parity anomaly in DNLSMs. This is precisely the question we address in this Letter for a $\mathcal{PT}$ symmetric DNLSM, where we use semiclassical Boltzmann equation approach to investigate the anomalous electrochemical transport, i.e. the Berry curvature current response induced by an electric field ${\bf{E}}$ and a gradient of a local chemical potential $\nabla \mu (\boldsymbol{r})$. In Weyl semimetals, the electrochemical conductivity tensor is nearly quantized (in terms of fundamental constants and the scattering time) when the Weyl nodes lie at different Fermi levels, being this a direct consequence of the chiral anomaly \cite{Rafael-Alberto-2021}. In a similar vein, recently nonuniform chemical potential profiles have been used to theoretically predict interesting phenomena in 2D and 3D materials. For example, they have been proposed as a mechanism to manipulate Majorana zero-modes in graphene armcharir nanoribbons \cite{PhysRevB.97.041414} and quantum wires situated in proximity to a $s$-wave superconductor \cite{PhysRevLett.105.177002, Alicea, PhysRevB.98.155414}. Furthermore, as shown in Ref. \cite{PhysRevB.95.235137} within the semiclassical Boltzmann formalism, gradients of chemical potential produce additional driving forces in the depletion region in a metal-semiconductor junction. Following these ideas, in this Letter we assume that nonuniform chemical potential profiles can be used to test the parity anomaly in $\mathcal{PT}$ symmetric DNLSMs.

In the relaxation time approximation, we derive a general formula for the nonlinear conductivity tensor $\sigma _{ijk} $, defined by $J_{i}=\sigma _{ijk} (\partial _{j} \mu ) E_{j}$. Introducing a small $\mathcal{PT}$-breaking mass term, we evaluate analytically the conductivity tensor, which nevertheless vanishes in the $\mathcal{PT}$-symmetric limit. We apply our results to the hexagonal pnictide CaAgP. In order to obtain a direct fingerprint of the parity anomaly, we introduce a tilting term which does not spoil the $\mathcal{PT}$ symmetry which guarantees the topological protection of the nodal ring. We show that for in-plane chemical gradient and electric field, an anomalous current is induced in the direction perpendicular to the nodal loop. Also, we verify that temperature effects do not destroy the signature of the parity anomaly.

\section{The model}

In the continuum approximation, the minimal Hamiltonian for a Dirac nodal-line semimetal with a single Dirac ring is given by \cite{Schnyder-2016}:
\begin{equation}
    H ({\boldsymbol{k}}) =  \frac{1}{\Lambda} \left( k _{0} ^{2} -  k _{\rho} ^{2} \right)   \sigma _{z} + v _{z} k _{z} \sigma _{y} + m \sigma _{x} \,  , \label{NLS-Hamiltonian}
\end{equation}
where the Pauli matrices $\sigma _{i}$ represent an effective orbital basis (not necessarily the spin degree of freedom) and $k _{\rho} ^{2} = k _{x} ^{2} + k _{y} ^{2}$. The parameters $\Lambda$ and $v _{z}$ arise from any particular lattice realization of the ${\boldsymbol{k}} \cdot {\boldsymbol{p}}$ model (\ref{NLS-Hamiltonian}). For later use we have introduced a small $\mathcal{PT}$-breaking mass $m \sigma _{x}$ which could be generated, for example, by inversion-breaking uniaxial strain, pressure, or an external electric field \cite{Du, Rendy}. The eigenenergies of the Hamiltonian are given by
\begin{equation}
    E _{s} ({\boldsymbol{k}}) = s \, \mathcal{E} ({\boldsymbol{k}})  \equiv s \sqrt{m ^{2} + v _{z} ^{2} k _{z} ^{2} + \frac{1}{\Lambda ^{2}} \left( k _{0} ^{2} -  k _{\rho} ^{2} \right) ^{2} } . \label{NLS-spectra}
\end{equation}
where $s= \pm 1$ is the band index. In the $\mathcal{PT}$-symmetric case, when $m$ is zero, the model consists in two bands that touch each other at a one-dimensional ring of radius $k_{0}$ in momentum space (located in the $k_{z}=0$ plane).

The symmetry protection of the Dirac ring (\ref{NLS-spectra}) is guaranteed by a quantized $\mathbb{Z}_{2}$ topological charge, which is given by the phase winding of the Bloch states around the nodal loop. Therefore, any $\mathcal{PT}$ symmetry-preserving perturbation cannot remove the nodal ring and open a gap. In the presence of spin-orbit coupling, several combinations of symmetries can stabilize the nodal lines \cite{PhysRevB.92.081201}. From a quantum field theory perspective, it is interesting that the $\mathbb{Z}_{2}$ invariant can be understood as a manifestation of the parity anomaly which occurs for fermions in (2+1) dimensions. This is possible because the low-energy excitations of the DNLSM, being a metallic system in (3+1) dimensions, can be described by a family of (2+1)-dimensional massless Dirac systems. The parity anomaly emerges since any gauge symmetric regularization of the quantum theory necessarily breaks $\mathcal{PT}$ symmetry. As a consequence, a parity-breaking Chern-Simons term is induced \cite{PhysRevD.29.2366, PhysRevLett.53.2449}, that gives  rise to a finite quantized anomalous Hall current for each (2+1) subsystem. However, when summed over the whole nodal ring, the total current vanishes. This is so because each point at the nodal line has a partner related by inversion symmetry which contributes with the opposite sign. Therefore, in the presence of an in-plane electric field, fermions at opposite sides of the nodal line flow to opposite directions (as in the valley Hall effect of graphene, where electrons from different valleys flow to opposite transverse edges). The only way to measure this effect is by filtering electrons by their momentum, as dumbbell filter devices do.

DNLSMs exhibit various anomalous transport responses due to the nontrivial structure of the Fermi surface, which includes linear \cite{PhysRevB.97.161113, PhysRevB.96.161105} and nonlinear \cite{PhysRevB.98.155125} Hall effects, the Kerr effect \cite{PhysRevB.104.125411, PhysRevB.103.165104} and giant nonlinear response in the presence of magnetic fields \cite{PhysRevB.104.245141}. All of these can be regarded as direct signatures of the $\mathbb{Z}_2$ topological invariant.

\section{Electrochemical transport in kinetic theory} 

We study the nonlinear electrochemical transport using the semiclassical formalism. In short, we investigate the current response of a DNLSM subjected to a spatially varying chemical potential $ \mu({\boldsymbol{r}})$ and an external homogeneous and static electric field ${\bf{E}}$. We start with the corresponding semiclassical equations of motion for an electron wave packet in a metal \cite{Xiao}:
\begin{equation}
    \dot{{\boldsymbol{r}}} _{s} = {\bf{v}} _{s} ({\boldsymbol{k}})  - \dot{{\boldsymbol{k}}} \times \boldsymbol{\Omega} _{s} ({\boldsymbol{k}}) , \quad \dot{{\boldsymbol{k}}} = - (e/ \hbar ) {\bf{E}} \label{SemiclassicalEqs} , 
\end{equation}
where ${\bf{v}} _{s} ({\boldsymbol{k}}) = \frac{1}{\hbar} {\nabla} _{{\boldsymbol{k}}} E _{s} ({\boldsymbol{k}})$ is the band velocity, ${\boldsymbol{k}}$ is the crystal momentum and $\boldsymbol{\Omega} _{s} ({\boldsymbol{k}}) = i \left< {\nabla} _{{\boldsymbol{k}}} u _{s} ({\boldsymbol{k}}) \vert \times \vert {\nabla} _{{\boldsymbol{k}}} u _{s} ({\boldsymbol{k}}) \right>$ is the Berry curvature defined in terms of the Bloch eigenstates $\left| u _{s} ({\boldsymbol{k}}) \right>$, i.e. $H ({\boldsymbol{k}}) \left| u _{s} ({\boldsymbol{k}}) \right> = E _{s} ({\boldsymbol{k}}) \left| u _{s} ({\boldsymbol{k}}) \right>$. The presence of the Berry curvature in the equations of motion (\ref{SemiclassicalEqs}) gives rise to anomalous transport perpendicular to the applied electric field. Evaluation of the Hall current from this term reproduces the Karplus-Luttinger formula for the anomalous Hall conductivity \cite{PhysRev.95.1154}.

The charge current is defined by
\begin{equation}
    {\bf{J}} = - e \sum _{s = \pm 1} \int \frac{d^{3} {\boldsymbol{k}} }{(2 \pi ) ^{3} } \, \left[ \dot{{\boldsymbol{r}}} _{s} - \frac{1}{e} \nabla _{{\boldsymbol{r}}} \times \boldsymbol{m} _{s} ({\boldsymbol{k}})  \right] \, f_{s} ({\boldsymbol{r}} , {\boldsymbol{k}} ) , \label{Current}
\end{equation}
where $f_{s} ({\boldsymbol{r}} , {\boldsymbol{k}})$ is the nonequilibrium quasiparticle distribution function which satisfies the Boltzmann equation. In the relaxation time approximation, the Boltzmann equation reads
\begin{equation}
    {\bf{v}} _{s} \cdot {\nabla} _{{\boldsymbol{r}}} \, f _{s} + \frac{e}{\hbar} {\bf{E}} \cdot ( \boldsymbol{\Omega}_{s} \times {\nabla} _{{\boldsymbol{r}}} - {\nabla} _{{\boldsymbol{k}}} ) \, f _{s} = - \frac{f _{s} - f _{s} ^{(0)}}{\tau} , \label{BoltzmanEq2}
\end{equation}
where we have omitted all the dependence on ${\boldsymbol{r}}$ and ${\boldsymbol{k}}$ for simplicity. The parameter $\tau$ is the transport time and $f _{s} ^{(0)}$ is the equilibrium Fermi-Dirac \textit{local} distribution defined by local temperature $T({\boldsymbol{r}})$ and local chemical potential $\mu ({\boldsymbol{r}})$. The second term in Eq. (\ref{Current}) is a contribution of the magnetization current, which is defined in terms of the orbital magnetic moment $\boldsymbol{m} _{s} ({\boldsymbol{k}})  = - i \frac{e}{2 \hbar} \left< {\nabla} _{{\boldsymbol{k}}} u _{s} ({\boldsymbol{k}}) \vert \times \left[ H ({\boldsymbol{k}}) - E _{s} ({\boldsymbol{k}}) \right] \vert {\nabla} _{{\boldsymbol{k}}} u _{s} ({\boldsymbol{k}}) \right> $, which generically describes the rotation of a wave packet around its center of mass \cite{PhysRevB.53.7010, PhysRevB.59.14915}.

Since we are interested in the nonlinear response, we recursively solve the Boltzmann equation (\ref{BoltzmanEq2}) assuming that $f _{s} = f _{s} ^{(0)} + f _{s} ^{(1)} + f _{s} ^{(2)}$, where $f _{s} ^{(1)} \sim \mathcal{O} ( E _{i} ) + \mathcal{O} ( \partial _{i} \mu)$ and  $f _{s} ^{(2)} \sim \mathcal{O} (E _{i} \, \partial _{j} \mu)$  contain the linear and the nonlinear terms, respectively. One finds
\begin{equation}
\label{DistributionFunc}
\begin{split}
f _{s} ^{(1)} &= \tau \, {\bf{v}} _{s} \cdot ( - e {\bf{E}} + {\nabla} _{{\boldsymbol{r}}} \, \mu ) \; \frac{\partial f _{s} ^{(0)}}{\partial E _{s}} , \\ f _{s} ^{(2)} &= \frac{e \tau}{\hbar} ( {\bf{E}} \times \boldsymbol{\Omega} _{s} \cdot {\nabla} _{{\boldsymbol{r}}} \, \mu ) \; \frac{\partial f _{s} ^{(0)}}{\partial E _{s}} . 
\end{split}
\end{equation}
It is worth mentioning that there are two more terms not listed in $f _{s} ^{(2)}$ which are independent of the Berry curvature. In a real experimental situation these could be relevant and therefore one should be able to distinguish between the topological and nontopological contributions. In the problem at hand, for $\mathcal{PT}$-symmetric DNLSMs, they do not contribute to the electrochemical current and hence in the following we will restrict ourselves to the analysis of the geometrical contribution. 

\section{Electrochemical transport in DNLSM}

\begin{figure}
    \centering
    \includegraphics[scale=0.7]{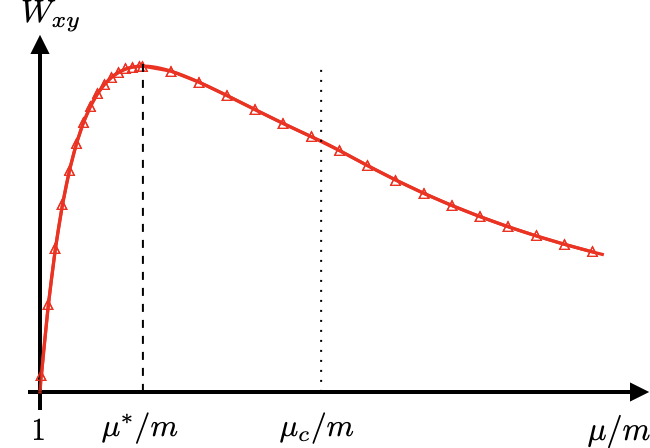}
    \caption{Plot of the function $W _{xy}$, given by Eq. (\ref{Fin_Wxy}), as a function of the (dimensionless) chemical potential $\mu / m$. The vertical dotted line  represents the critical value $\mu _{c}/ m$, while the vertical dashed line corresponds to the critical value $\mu ^{\ast}/ m$ which maximizes the function. } \label{Wxy}
\end{figure}

After computing all the relevant components of the nonequilibrium distribution function $f _{s}$, one then plugs it and the velocity $\dot{{\boldsymbol{r}}} _{s}$ defined in Eq. (\ref{SemiclassicalEqs}) into the quasiparticle current (\ref{Current}) to get the electrochemical current. For a DNLSM as modelled by the Hamiltonian (\ref{NLS-Hamiltonian}) the orbital magnetic moment and the Berry curvature are related by $\boldsymbol{m} _{s} ({\boldsymbol{k}}) = - s \frac{e}{\hbar} \mathcal{E} ({\boldsymbol{k}}) \, \boldsymbol{\Omega} _{s} ({\boldsymbol{k}})$. All in all, the nonlinear current response can be expressed as $J_{i}=\sigma_{ijk}  (\partial _{j} \mu ) E_{k}$ (latin indices span the cartesian components $\{\ \!\! x, y, z \}\ \!\!$), where the nonlinear conductivity tensor $\sigma_{ijk}$ is given by
\begin{equation}
    \sigma_{ijk} (\mu ) = \sigma _{0} \, \left[ \epsilon _{lik} W_{jl} (\mu ) + \epsilon _{ljk} W_{il} (\mu ) + \epsilon _{lij} M_{kl} (\mu ) \right] , \label{Nonlinear-Cond-Tensor}
\end{equation}
where $\sigma _{0} \equiv e^{2}\tau / h ^{2}$ is the conductance quantum and we have defined the dimensionless integrals
\begin{equation}
\label{Functions_W_M}
\begin{split}
W_{ij} (\mu ) &= - \frac{\hbar}{2 \pi} \sum _{s = \pm 1} \int v _{si} ({\boldsymbol{k}}) \Omega _{sj} ({\boldsymbol{k}}) \frac{\partial f ^{(0)} _{s} ({\boldsymbol{k}}) }{\partial E_{s}}  d ^{3} {\boldsymbol{k}} ,  \\ M_{ij} (\mu ) &= - \frac{\hbar}{2 \pi} \sum _{s = \pm 1} \int v _{si} ({\boldsymbol{k}}) \Omega _{sj} ({\boldsymbol{k}}) \mathcal{E} ({\boldsymbol{k}}) \frac{\partial ^{2} f ^{(0)} _{s} ({\boldsymbol{k}}) }{\partial E_{s} ^{2}}  d ^{3} {\boldsymbol{k}} .
\end{split}
\end{equation}
Physically, the function $W_{ij}$ is determined by the anomalous velocity (\ref{SemiclassicalEqs}) and the second-order correction to the quansiparticle distribution (\ref{DistributionFunc}), whilst the function $M_{ij}$ arises from the magnetization current in Eq. (\ref{Current}). We now focus on the evaluation of the above integrals for the DNLSM. The geometry of the nodal line advises the use of the cylindrical coordinates $\{\ \!\! k_{\rho},\varphi,k_{z} \}\ \!\!$. The band velocity and Berry curvature for the model Hamiltonian (\ref{NLS-Hamiltonian}) are
\begin{align}
    {\bf{v}} _{s} ({\boldsymbol{k}}) = s \frac{2 k _{\rho} (k _{\rho} ^{2} - k _{0} ^{2})}{\hbar \Lambda ^{2} \mathcal{E} ({\boldsymbol{k}}) } \hat{\boldsymbol{e}} _{\rho} + s \frac{v_{z} ^{2} k _{z}}{\hbar \mathcal{E} ({\boldsymbol{k}}) } \hat{\boldsymbol{e}} _{z} \label{Velocity}
\end{align}
and
\begin{equation}
    \boldsymbol{\Omega} _{s} ({\boldsymbol{k}}) = s \frac{m v _{z} k _{\rho}}{\Lambda \mathcal{E} ^{3} ({\boldsymbol{k}}) } \hat{\boldsymbol{e}} _{\varphi} ,  \label{Berry-Curvature}
\end{equation}
respectively. Here, $ \hat{\boldsymbol{e}} _{\rho} = \cos \varphi \, \hat{\boldsymbol{e}} _{x} + \sin \varphi  \, \hat{\boldsymbol{e}} _{y}$ and $\hat{\boldsymbol{e}} _{\varphi} = \hat{\boldsymbol{e}} _{z} \times \hat{\boldsymbol{e}} _{\rho} $. Besides, we will work at zero temperature, so the derivative of the equilibrium distribution function $f ^{(0)} _{s}$ will be strongly peaked at the Fermi level $\mu$. This means that the nonlinear transport coefficients are properties of the Fermi surface. Without loss of generality, we will assume that the Fermi level $\mu$ crosses the conduction band, i.e. $\mu >m$. The axial symmetry of the nodal line implies that the only nonzero components of the functions (\ref{Functions_W_M}) are $W_{xy} = - W_{yx}$ and $M_{xy} = - M_{yx}$. Substituting the velocity (\ref{Velocity}) and Berry curvature (\ref{Berry-Curvature}) into Eqs. (\ref{Functions_W_M}) we obtain
\begin{equation}
    W _{xy} (\mu) = \frac{m v _{z}}{\Lambda ^{3} \mu ^{4}} \int _{- \infty} ^{+ \infty} \!\! \int _{0} ^{\infty} \!\! k _{\rho} ^{3} (k _{\rho} ^{2} - k _{0} ^{2}) \,  \delta ( \mu - \mathcal{E} ) \,  d k _{\rho} dk _{z} ,  \label{wxy}
\end{equation}
and for the function $M_{yx}$ one can spot the identity $M_{xy} = - \frac{\partial}{\partial \mu} (\mu W_{xy})$. To evaluate the integral in Eq. (\ref{wxy}) we use the properties of the Dirac delta in composition with an arbitrary continuously differentiable function $f(x)$, i.e. $\delta (f(x))= \sum _{i} \frac{\delta (x-x_{i})}{\vert f ^{\prime} (x_{i}) \vert }$, where the sum extends over all roots $x_{i}$ of the function $f(x)$. In the present case we have $\mu = \mathcal{E} (k _{\rho}, k _{z})$, wherefrom we obtain the roots
\begin{equation}
    k _{z \pm} = \pm k _{z} ^{\ast} \equiv \pm \frac{1}{v _{z}} \sqrt{ (\mu ^{2} - m ^{2})  - \frac{1}{\Lambda ^{2}} \left( k _{0} ^{2} - k _{\rho}  ^{2} \right) ^{2}  } ,
\end{equation}
such that $\delta ( \mu - \mathcal{E}  ) = \frac{\mu}{v _{z} ^{2} k _{z} ^{\ast}} \big[ \delta (k _{z} - k _{z} ^{\ast}) + \delta (k _{z} + k _{z} ^{\ast}) \big] \Theta (k _{z} ^{\ast 2})$, where the Heaviside function $\Theta$ guarantees that the root $k _{z} ^{\ast}$ is real valued. Inserting this result into the integral expression (\ref{wxy}) and integrating with respect to $k_{z}$ we obtain
\begin{equation}
    W _{xy} (\mu) = \frac{2m}{ \Lambda ^{3} \mu ^{3}} \int _{0} ^{\infty} \frac{k _{\rho} ^{3} (k _{\rho} ^{2} - k _{0} ^{2}) \, \Theta (k _{z} ^{\ast  2})}{\sqrt{ (\mu ^{2} - m ^{2})  - \frac{1}{\Lambda ^{2}} \left( k _{0} ^{2} - k _{\rho}  ^{2} \right) ^{2}  }} d k _{\rho} .  \label{Wxy-2}
\end{equation}
The constraint imposed by the step function $\Theta (k _{z} ^{\ast  2})$ defines the limits of integration over the variable $k _{\rho}$. In fact, it defines the region
\begin{equation}
    \mbox{Re} \left[ \sqrt{k _{0} ^{2} - \Lambda \sqrt{ \mu ^{2} - m ^{2}}  } \, \right] \leq k _{\rho} \leq  \sqrt{k _{0} ^{2} + \Lambda \sqrt{ \mu ^{2} - m ^{2}}  } ,  \label{Condition}
\end{equation}
which motivates the introduction of a critical Fermi level $\mu _{c} = \sqrt{m^{2} + (k_{0}^{2}/ 2 \Lambda) ^{2}}$ and the radii
\begin{equation}
    k _{\rho \pm} ^{\ast} = \sqrt{\Lambda \left(  \sqrt{ \mu _{c} ^{2} - m ^{2}} \pm  \sqrt{ \mu ^{2} - m ^{2}} \right) } ,
\end{equation}
such that the condition (\ref{Condition}) renders: (I) $k_{\rho} \in [0 , k _{\rho +} ^{\ast}]$ for $\mu > \mu _{c}$ and (II) $k_{\rho} \in [k _{\rho -} ^{\ast} , k _{\rho +} ^{\ast}]$ for $\mu < \mu _{c}$. Now we are ready to perform the integral (\ref{Wxy-2}). Changing variable to $x = k_{\rho}/k_{0}$, Eq. (\ref{Wxy-2}) becomes
\begin{equation}
\begin{split}
W _{xy} (\mu ) &= \frac{2m k _{0} ^{6}}{ \Lambda ^{3} \mu ^{3}} \left[ \Theta (\mu _{c} - \mu ) \int _{x _{-}} ^{x _{+}}  + \Theta (\mu - \mu _{c} ) \int _{0} ^{x _{+}}  \right]  \\  & \times \frac{x ^{3} (x ^{2} - 1) \, d x  }{\sqrt{ (\mu ^{2} - m ^{2})  -(\mu _{c} ^{2} - m ^{2})  \left( 1 - x ^{2} \right) ^{2}  }}, 
\end{split}
\end{equation}
where $x_{\pm} \equiv k _{\rho \pm} ^{\ast} / k _{0}$. This integral can be evaluated in an analytical fashion. Defining the ratio $ \gamma \equiv \frac{\mu ^{2} - m ^{2}}{\mu _{c} ^{2} - m ^{2}} $ we finally obtain
\begin{equation}
\label{Fin_Wxy}
\begin{split}
W _{xy} (\mu) &= \frac{\pi m}{2 \mu} \left[ 1 - \left( \frac{m}{\mu} \right) ^{2} \right] \bigg\{\ \!  \Theta (\mu _{c} - \mu ) \, + \, \Theta (\mu - \mu _{c} ) \\  & \times \left[ \frac{1}{2}  + \frac{\sqrt{\gamma -1}}{ \pi \gamma} + \frac{1}{\pi} \mbox{arccot} \left( \sqrt{\gamma -1} \right)  \right]  \bigg\}\ ,
\end{split}
\end{equation}
which fully determines the nonlinear conductivity tensor (\ref{Nonlinear-Cond-Tensor}). In Fig. \ref{Wxy} we plot $W _{xy}$ as a function of the (dimensionless) chemical potential $\mu / m$. The vertical dotted line marks the critical value $\mu _{c}/ m$ and the vertical dashed line corresponds to the critical value $\mu ^{\ast}/ m$ which maximizes the function $W _{xy}$. We observe that this function vanishes for $\mu = m$, increases monotonically up to a maximum located at $\mu ^{\ast}= \sqrt{3} m < \mu _{c}$ (for which $W _{xy} (\mu ^{\ast})=\pi / 3 \sqrt{3}$) and then decreases monotonically for $\mu > \mu ^{\ast}$. Note that it is continuous at the critical Fermi level $\mu = \mu _{c}> \mu ^{\ast}$.

To discuss the electrochemical response in a realistic DNLSM, it is convenient to consider the precise values of the parameters appearing in the low-energy model (\ref{NLS-Hamiltonian}). From a lattice model for the hexagonal pnictide CaAgP, ab-initio calculations give $\Lambda = - 0.083 \textup{\r{A}}^{-2} \mbox{eV} ^{-1}$, $k_{0} = 0.207 \textup{\r{A}} ^{-1}$ and $v_{z} = - 2.2 \textup{\r{A}}$eV \cite{doi:10.7566/JPSJ.85.123701, PhysRevB.95.245113}. For $m=0.05$eV, the critical Fermi energy is found to be $\mu _{c} \approx 0.94$eV. However, the Dirac nodal line is robust against weak short-range correlated disorder for moderate chemical potentials ($\mu < 0.5$eV). So, only the region $\mu < \mu _{c}$ in Eq. (\ref{Fin_Wxy}) would be relevant in a realistic situation.

\begin{figure}
    \centering
    \includegraphics[scale=0.5]{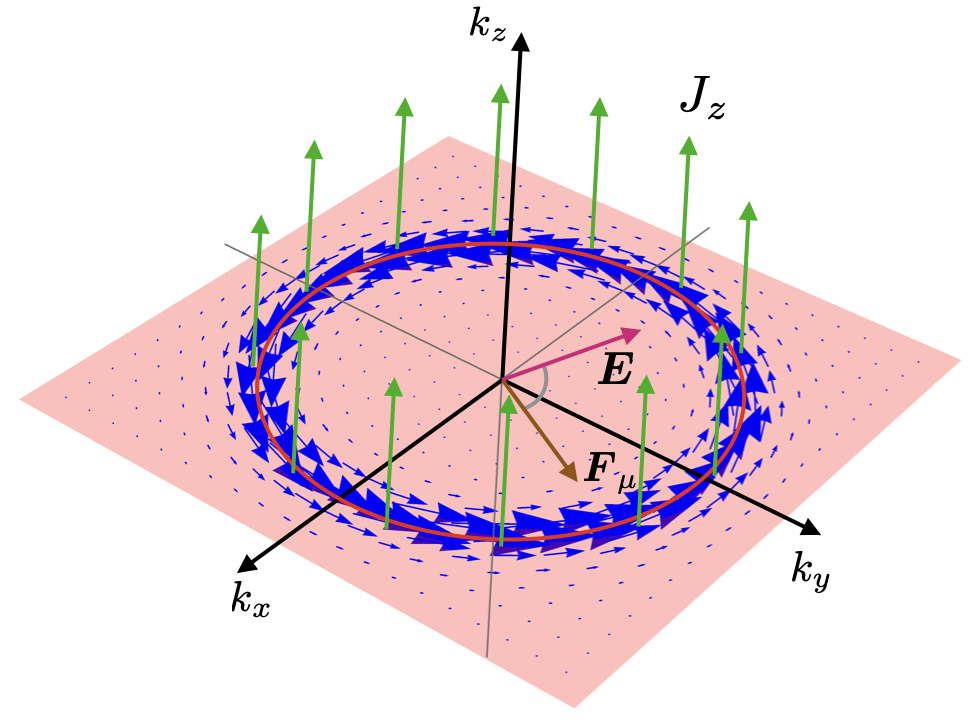}
    \caption{The blue arrows indicate the Berry curvature, given by Eq. (\ref{Berry-Curvature}), and the red circle marks the nodal ring. The in-plane electric and electrochemical fields are represented by the pink and brown arrows, respectively. The green arrows represent the topological electrochemical current.}    \label{CurrentPlot}
\end{figure}

The electrochemical current $J_{i}=\sigma_{ijk} (\partial _{j}\mu ) E_{k}$ is an experimentally observable signature of the nontrivial topology of the DNLSM. Figure \ref{CurrentPlot} schematically illustrates the Berry curvature (blue arrows) and the plane $k_{z}=0$. By choosing in-plane chemical potential gradient $ \nabla \mu = \mathcal{\boldsymbol{F}}_{\mu} $ (brown arrow) and electric field $\boldsymbol{E}$ (pink arrow), i.e. $\mathcal{\boldsymbol{F}}_{\mu} \cdot \hat{\boldsymbol{e}} _{z} = \boldsymbol{E} \cdot \hat{\boldsymbol{e}} _{z} = 0$, the local electrochemical current points along the $z$ direction (green arrow). We get
\begin{equation}
    J_{z} (\mu ) = \sigma _{0} \frac{\pi m}{2 \mu} \left[ 1 - 3 \left( \frac{m}{\mu} \right) ^{2} \right]   \, \mathcal{\boldsymbol{F}}_{\mu} \cdot \boldsymbol{E}  .  \label{ElectroChemicalCurrent}
\end{equation}
We emphasize that for the $\mathcal{PT}$-symmetric DNLSM considered in this work, the nontopological contributions to the electrochemical current vanish by symmetry. However, this may not be true for general DNLSMs protected by other symmetries such as crystalline and time-reversal.The current response (\ref{ElectroChemicalCurrent}) is quite interesting. As discussed in Ref. \cite{PhysRevB.97.161113}, the anomalous Hall current vanishes for a DNLSM. The reason behind this cancellation is that the nodal ring can be foliated into a family of $(2+1)$-dimensional subsystems parametrized by the polar angle $\varphi$ in momentum space, such that each point along the nodal line has another point related by inversion symmetry which contributes oppositely to the current, so the total transverse current vanishes. The only possibility to measure the anomalous Hall current is by means of a dumbbell device which filters electrons by their momentum. In this sense, the predicted electrochemical current (\ref{ElectroChemicalCurrent}) can be clearly distinguished from the anomalous Hall effect and it can be measured with simple contacts.

It is interesting that from the field theoretical point of view, the anomalous Hall effect is a fingerprint of the parity anomaly in DNLSMs, i.e. the Hall current survives to the limit $m \to 0$ \cite{PhysRevB.97.161113}. As pointed out in Ref. \cite{PhysRevB.98.155125}, the nonlinear response of DNLSMs is also related to the parity anomaly. In the present case, Fig. \ref{Wxy} reveals that the current response vanishes in the limit $m \to 0$. Therefore, the electrochemical current (\ref{ElectroChemicalCurrent}) is nonzero only for $\mathcal{PT}$ symmetry-breaking DNLSMs, which is quite realistic since the mass term can be induced by inversion-breaking uniaxial strain, pressure, or an external electric field \cite{Du, Rendy}. It is natural to ask whether is possible to obtain an imprint of the parity anomaly in the electrochemical transport in DNLSMs, i.e. in the $\mathcal{PT}$ symmetric case ($m \to 0$). We answer this question in the affirmative in the next Section.
 
\section{Parity anomaly induced transport} \label{ParitySec}

As showed in the previous section, the electrochemical current vanishes in the limit $m \to 0$, so, it cannot be regarded as a signature of the parity anomaly in DNLSMs. In order to search for a direct signature of the anomaly, it is mandatory to consider a $\mathcal{PT}$ symmetric model, since the stability of the nodal line comes from the invariance of the combined symmetry. This suggests that a tilting term of the form $\hbar {\boldsymbol{\upsilon}} \cdot {\boldsymbol{k}} \sigma _{0}$ will do the work. Clearly, this term breaks separately $\mathcal{P}$ and $\mathcal{T}$, but does not spoil the combined $\mathcal{PT}$ symmetry of the model. Tilted nodal loops have been predicted to occur in alkaline-earth stannides, germanides, and silicides \cite{PhysRevB.93.201114} and in other materials displaying non-symmorfic symmetries \cite{PhysRevB.96.115106, Ekahana_2017}. All in all, let us consider the $\mathcal{PT}$-symmetric model
\begin{equation}
    H ({\boldsymbol{k}}) = \hbar {\boldsymbol{\upsilon}} \cdot {\boldsymbol{k}} \sigma _{0} + \frac{1}{\Lambda} \left( k _{0} ^{2} -  k _{\rho} ^{2} \right) \sigma _{z} + v _{z} k _{z} \sigma _{y} \,  , \label{PT-NLS-Hamiltonian}
\end{equation}
where ${\boldsymbol{\upsilon}}$ is the tilting velocity. The corresponding energy dispersion is $E_{s} = \hbar {\boldsymbol{\upsilon}} \cdot {\boldsymbol{k}} + s \sqrt{v _{z} ^{2} k _{z} ^{2} + \frac{1}{\Lambda ^{2}} \left( k _{0} ^{2} -  k _{\rho} ^{2} \right) ^{2} }$. In the following, we consider tilt only in the in-plane direction, because it produces electron and hole pockets. Without loss of generality we set ${\boldsymbol{\upsilon}} = \upsilon \, (\cos \alpha \, \hat{{\boldsymbol{e}}} _{x} + \sin \alpha \, \hat{{\boldsymbol{e}}} _{y})$ such that ${\boldsymbol{\upsilon}} \cdot {\boldsymbol{k}} = \upsilon \, k _{\rho} \cos ( \varphi  - \alpha )$, where $\varphi$ is the polar angle that defines the nodal line in momentum space. In the $m=0$ limit, the Berry curvature (\ref{Berry-Curvature}) is zero in the entire Brillouin zone, except at the nodal ring, where it becomes singular, i.e.  $\boldsymbol{\Omega} _{s} ({\boldsymbol{k}}) = s \pi   \delta (k _{z}) \delta (k_{\rho} - k _{0}) \hat{\boldsymbol{e}} _{\varphi}$ while the band velocity along the nodal ring is simply ${\bf{v}} _{s} = {\boldsymbol{\upsilon}}$. Substituting these expressions into Eq. (\ref{Functions_W_M}) we get
\begin{equation}
W_{ij} (\mu ) = \hbar \upsilon _{i} k _{0} \int _{0} ^{2 \pi } d \varphi  \, \hat{\boldsymbol{e}} _{\varphi j} \, \delta (\mu - \hbar \upsilon  k _{0} \cos ( \varphi - \alpha ) )  .  \label{Int}
\end{equation}
where we have used that the zero-temperature equilibrium distribution functions for electrons and holes become $f ^{(0)} _{s}= \Theta \{\ \!\! s [ \mu - \hbar \upsilon  k _{0} \cos ( \varphi - \alpha )  ] \}\ \!\! $ \cite{PhysRevB.98.155125}. To evaluate the integral (\ref{Int}) we use the properties of the Dirac delta in composition with an arbitrary continuously differentiable function. The argument of the delta function determines the angles $\varphi _{\chi} = \alpha + \chi \arccos \left( \mu / \mu _{c} \right) $, where $\chi = \pm 1$ and $\mu _{c} \equiv \hbar \upsilon k _{0}$ is a critical Fermi level. So Eq. (\ref{Int}) becomes


\begin{equation}
W_{ij} (\mu ) = \frac{\hat{\upsilon} _{i}}{\sqrt{1 - \left( \mu / \mu _{c} \right) ^{2} }} \sum _{\chi = \pm 1} \int _{0} ^{2 \pi } d \varphi  \, \hat{\boldsymbol{e}} _{\varphi j} \, \delta ( \varphi  - \varphi _{\chi} )  , 
\end{equation}
for $\mu< \mu_{c}$ and vanishes for $\mu> \mu_{c}$. Here  $\hat{\boldsymbol{\upsilon}} = \boldsymbol{\upsilon} / \upsilon $ is the unit vector along the tilting. This integral is quite simple. The final result is
\begin{equation}
W_{ij} (\mu ) = Q_{0} (\mu / \mu _{c}) \, \hat{\upsilon} _{i} (\hat{\boldsymbol{e}} _{z} \times \hat{\boldsymbol{\upsilon}} ) _{j} . \label{Q0-funct}
\end{equation}
where $Q_{0} (y)= \frac{2 y}{\sqrt{1-y^{2}}}$. For in-plane chemical potential gradient and electric field, the electrochemical current in the direction perpendicular to the nodal loop becomes
\begin{equation}
    J_{z} (\mu ) = - \sigma _{0} \frac{2 (\mu / \mu _{c})}{ \left[ 1 - (\mu / \mu _{c}) ^{2} \right] ^{3/2} }  \, ( \hat{\boldsymbol{\upsilon}} \cdot \mathcal{\boldsymbol{F}}_{\mu} )( \hat{\boldsymbol{\upsilon}} \cdot \boldsymbol{E} ) . \label{Anomalous2}
\end{equation}
This is a direct fingerprint of the parity anomaly in the nonlinear response of DNLSMs. Assuming a tilting of the order of the Fermi velocity, which is typically $v_{F} \approx 10 ^{8}$cm/s, and $k_{0} = 0.207 \textup{\r{A}} ^{-1}$ as for CaAgP, we find that $\mu_{c} \sim 1$eV. Since the nodal ring is robust against weak disorder for moderate chemical potential ($\mu < 0.5$eV), hence our predictions would been well-testable for an appropriate range of chemical potentials.

\section{Summary and discussion}

We have studied the electrochemical transport in Dirac nodal-line semimetals in the absence of spin-orbit coupling, where the combined $\mathcal{PT}$ symmetry stabilizes the Dirac ring. We first introduce a small $\mathcal{PT}$-breaking mass term, which could be induced by inversion-breaking uniaxial strain, pressure, or an external electric field \cite{Du, Rendy}. In this case, we compute the electrochemical current and find, for in-plane chemical potential gradient $\mathcal{\boldsymbol{F}}_{\mu}$ and electric field $\mathcal{\boldsymbol{E}}$, an anomalous current in the direction perpendicular to the Dirac ring when the Fermi level crosses the conduction band, i.e. $J_{z} \sim \mathcal{\boldsymbol{F}}_{\mu} \cdot \mathcal{\boldsymbol{E}}$. It is interesting the emergence of a critical Fermi level, which reflects in the nonlinear conductivity tensor (see Fig. \ref{Wxy}). Interestingly, when applied to the hexagonal pnictide CaAgP, the critical Fermi level is large as compared with the range of chemical potentials for which the linear response is stable under moderate perturbations. One can also observe the existence of a critical Fermi level at $\sqrt{3}m$ which maximizes the anomalous current. Although interesting, this result cannot be regarded as a direct consequence of the parity anomaly, since the current response vanishes in the $m=0$ limit.

In pursuit of a direct signature of the parity anomaly, we consider the $\mathcal{PT}$ symmetric model, given by Eq. (\ref{NLS-Hamiltonian}) with $m=0$, supplemented by a tilting term of the form  which does not spoil the symmetry. For in-plane tilting both the conduction and valence bands contribute to the transport. We find that for in-plane chemical potential gradient $\mathcal{\boldsymbol{F}}_{\mu}$ and electric field $\boldsymbol{E}$, a finite anomalous current in the direction perpendicular to the Dirac ring is induced, i.e. $J_{z} \sim ( \hat{\boldsymbol{\upsilon}} \cdot \mathcal{\boldsymbol{F}}_{\mu} )( \hat{\boldsymbol{\upsilon}} \cdot \boldsymbol{E} )$.

\begin{figure}
    \centering
    \includegraphics[scale=0.7]{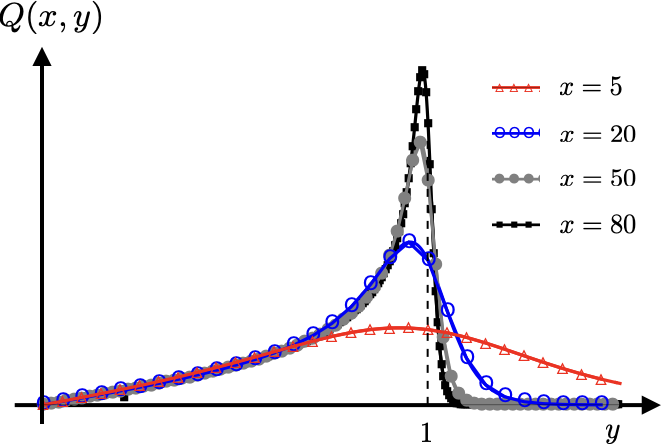}
    \caption{Plot of the function $Q (x,y)$, given by Eq. (\ref{Q-func}), as a function of the (dimensionless) chemical potential $y=\mu / \mu _{c}$ for different values of $x=\mu _{c}/ k _{B}T$. The vertical dashed line  marks the value $y=1$. } \label{PlotQ-func}
\end{figure}

One may wonder whether the signature of the parity anomaly appears only at zero temperature, or it takes place also for nonzero temperatures. We answer this question in the affirmative: the parity anomaly induced electrochemical transport appears also for $T>0$. This can be seen by generalizing the discussion of the previous section for nonzero temperatures. In this case, the function (\ref{Q0-funct}) generalizes to $W_{ij} (\mu , T) = Q ( \mu _{c} / k _{B} T, \mu / \mu _{c}) \, \hat{\upsilon} _{i} (\hat{\boldsymbol{e}} _{z} \times \hat{\boldsymbol{\upsilon}} ) _{j} $, where we have defined the dimensionless function
\begin{equation}
    Q (x,y) = y \frac{\partial}{\partial y} \int _{0} ^{2 \pi} \frac{1}{1 + e ^{x (\cos \phi - y )}} d \phi . \label{Q-func}
\end{equation}
In Fig. \ref{PlotQ-func} we plot $Q (x,y)$ as a function of the (dimensionless) chemical potential $y = \mu / \mu _{c}$ for different values of $x=\mu _{c} / k _{B}T$. The vertical dashed line marks the value $y=1$. We observe that this function vanishes at $y=0$. For a finite value of $x$, it increases monotonically up to a maximum value and then decreases asymptotically to zero for $y \to \infty$. Interestingly, as the temperature decreases (i.e. increasing $x$), the function $Q (x,y)$ confines progressively to the region $y<1$. In the zero-temperature limit (i.e. $x \to \infty$), the function (\ref{Q-func}) reduces exactly to the function $Q_{0}(y)$ which defines the coefficient appearing in Eq. (\ref{Q0-funct}), i.e. $\lim \limits _{x \to \infty} Q (x,y) \to Q_{0}(y) $, and collapses to zero for $y>1$, as expected. The two most peaked curves shown in Fig. \ref{PlotQ-func}, the gray and black lines, correspond to the temperatures $T \approx 230$K and $T \approx 140$K, respectively. All in all, the parity anomaly manifests in the electrochemical transport for both zero and nonzero temperatures.

Although experimental realizations of 3D Dirac materials are rather few and recent, one can establish the best candidate materials to test our predictions. The hexagonal pnictides CaAgP, CaAgAs and Ca$_3$P$_2$ are particularly promising, because they are available in single crystal form, possess a single nodal ring near the Fermi energy protected by the combined $\mathcal{PT}$ symmetry and the spin-orbit coupling is very weak. In particular, numerical estimates along this Letter have been done for the optimal  nodal-line CaAgP.

Finally, a precise control of the local chemical potential would be of great importance for realizing the topologically protected electrochemical current we predict. As discussed above, from the theoretical side, nonuniform chemical potential profiles have been proposed as a mechanism to manipulate Majorana fermions in graphene \cite{PhysRevB.97.041414} and quantum wires in proximity to a superconductor \cite{PhysRevLett.105.177002, Alicea, PhysRevB.98.155414}. In Weyl semimetals, for example, it has been theoretically predicted that the local chemical potential can be controlled by doping or having inhomogeneous impurities along the sample  \cite{PhysRevB.98.081114, PhysRevB.91.195107}. Surprisingly, the local control of the chemical potential has been routinely implemented in experiments by an array of gates along the sample. For example, in Refs. \cite{doi:10.1126/science.1222360, doi:10.1126/sciadv.1701476}, this method has been used to find signatures of Majorana fermions in superconductor-semiconductor nanowire devices. Also, as experimentally shown in Ref. \cite{doi:10.1021/acsaelm.0c00701} for the topological insulator Bi$_{2-x}$Sb$_x$Se$_3$, the active manipulation of trapped charges offers a powerful way to engineer spatial distributions of the chemical potential (significantly more efficient than conventional gating), for instance, by creating n$^{+}-$n and p$-$n junctions along the nanoribbon devices. In closing, as far as we know, no experiments have  been yet carried out in nodal-lines by using nonuniform profiles of the chemical potential, but undoubtedly, it will be achievable in the near future.

\acknowledgments
L.M.O. was supported by the CONACyT PhD fellowship No. 834773. A.M.R has been partially supported by DGAPA-UNAM Project No. IA102722 and by Project CONACyT (M\'{e}xico) No. 428214.

\bibliographystyle{main}

\bibliography{EM-NLSM}

\end{document}